\documentclass[aps,prl,showpacs,floatfix,twocolumn]{revtex4}
\usepackage{epsfig}
\usepackage{graphicx}
\usepackage{amsmath}
\usepackage{hhline}
\usepackage{amssymb}
\usepackage{times}

\newlength{\dinwidth}
\newlength{\dinmargin}
\begin{document}
\newcommand{\be}{\begin{equation}}
\newcommand{\ee}{\end{equation}}
\newcommand{\bea}{\begin{eqnarray}}
\newcommand{\eea}{\end{eqnarray}}
\title{Probing small $x$ parton densities in ultraperipheral $AA$ and 
\\$pA$ collisions at the LHC}
% Force line breaks with
%\\
\author{Mark Strikman}
 \email{strikman@phys.psu.edu}
 \affiliation{Pennsylvania State University, University Park, PA 16802, USA}
\author{Ramona  Vogt}
 \email{vogt@lbl.gov  }
\affiliation{Department of Physics, University of California, Davis, CA 95616,
USA \\
 and Nuclear Science Division LBNL, Berkeley, CA 94720, USA}
\author{Sebastian White}
 \email{white1@bnl.gov }
\affiliation{Department of Physics, Brookhaven National Laboratory, 
Upton, NY 11973, USA}
\date{\today}
\begin{abstract}
%\noindent 
We calculate production rates for several hard processes in
ultraperipheral proton-nucleus and nucleus-nucleus collisions at the LHC. 
The resulting high rates demonstrate that some key directions
in small $x$ research proposed for HERA will be accessible at the LHC through
these ultraperipheral processes.
Indeed, these measurements can extend the HERA $x$ range by roughly a factor 
of 10 for similar virtualities.  Nonlinear effects
on the parton densities will thus be significantly more important in these
collisions than at HERA.

\end{abstract}
\maketitle

Studies of small $x$ deep inelastic scattering at HERA substantially 
improved our understanding of  strong interactions at high energies.
Among the key findings of HERA were the direct observation of the rapid growth 
of the small $x$ structure functions over a wide range of virtualities,
$Q^2$, and the observation of a significant probability for hard diffraction
consistent with approximate scaling and a logarithmic $Q^2$ dependence 
(``leading twist" dominance).
HERA also established a new class of hard exclusive processes -- high $Q^2$
vector meson production -- described by the QCD factorization theorem and 
related to generalized parton distributions in nucleons.

The importance of 
nonlinear QCD dynamics at small $x$ is one of the focal points of 
theoretical activity (see {\it e.g.}\ Ref.~\cite{Mueller}).
Analyses suggest that the strength of the interactions, especially when
a hard probe directly couples to gluons, approaches the
maximum possible strength -- the black disk limit -- for $Q^2 \le 4$ GeV$^2$.
These values are relatively small, with an even smaller $Q^2$
for coupling to quarks, $Q^2 \sim  1\,$ GeV$^2$, 
making it difficult to separate 
perturbative and  nonperturbative effects at small $x$ and $Q^2$. 
Possible new directions for further experimental investigation of this regime
include higher energies, nuclear beams and studies of the longitudinal virtual 
photon cross section, $\sigma_L$. The latter two options were discussed for 
HERA \cite{Arneodo:1996qa,Alexopoulos:2003aa}. Unfortunately, it now seems 
that HERA will stop operating in two years with no further measurements along 
these lines except perhaps of $\sigma_L$.
One might therefore expect that experimental investigations in this
direction would end during the next decade.

The purpose of this letter is to demonstrate that several of the crucial 
directions of HERA research can be continued and extended by studies of 
ultraperipheral heavy ion collisions (UPCs) at the LHC.
UPCs are interactions of two heavy nuclei (or a proton and a nucleus) in which 
a nucleus emits a quasi-real photon that interacts 
with the other nucleus (or proton).  These collisions have the distinct
feature that the photon-emitting nucleus either does not break up or only 
emits a few neutrons through Coulomb excitation, leaving a substantial 
rapidity gap in the same direction.  These kinematics can be readily 
identified by the hermetic LHC detectors, ATLAS and CMS.
In this paper we consider the feasibility of studies in two of the directions 
pioneered at HERA: parton densities and hard diffraction. The third, quarkonium production, was discussed previously 
\cite{Lippmaa:1997qb,Bertulani:2005ru,pareport}. 
It was shown that $pA$ and $AA$ scattering can extend the energy range
of HERA, characterized by $\sqrt{s_{\gamma N}}$, by about a factor of 10 and,
in particular, investigate
the onset of color opacity for quarkonium photoproduction.

 %--------------------------------------------
\begin{figure}[tbh]
%\vspace{9pt}
\includegraphics[width=6cm]{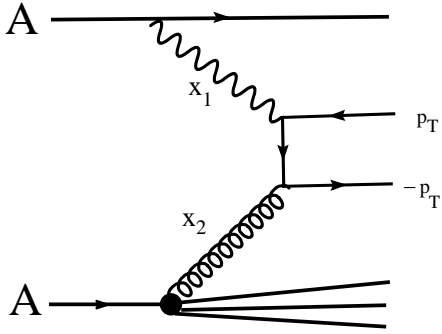}
\caption{Diagram of dijet production by photon-gluon fusion
where the photon carries momentum fraction $x_1$ while the
gluon carries momentum fraction $x_2$.}
\label{1}
\end{figure}
%--------------------------------------------

Though we can only study reactions initiated by a quasi-real photon, this is 
not a disadvantage in the study of new phenomena relative to HERA. 
Indeed, nonlinear effects should set in earlier 
in the gluon sector, more easily accessed through photon-gluon interactions 
than inclusive electron-hadron scattering.
In the following discussion, we assume that DGLAP evolution of the parton
densities holds in the region we explore. For definiteness, we discuss 
measurements of the nucleus (proton) inclusive parton distribution functions 
(PDFs) and diffractive PDFs, even though it appears that part of the 
accessible kinematics is within the domain where the DGLAP 
approximation is likely to break down.
In principle, the nuclear (proton) PDFs (though not the diffractive PDFs) 
could be studied  in $pp$ and $pA$ collisions. 
However, PDF studies in the $p_T$ range we explore are impossible in hadronic
interactions due to the large background from {\it e.g.}\ multiple jet 
production.  
Photon-nucleus interactions are thus much cleaner than proton-nucleus 
interactions. In addition, the LHC $pA$ program is likely to begin several 
years after the start of the heavy ion program.

In our study, we calculate dijet production to leading order (LO)
in nucleus-nucleus collisions, as in Ref.~\cite{ramona}. 
Based on HERA studies, we use 
a 5 GeV $p_T$ cutoff for the applicability of perturbative QCD to jet
production.  We also assume the ATLAS detector coverage and performance 
as discussed below.
We use the MRST LO nucleon PDFs \cite{MRST} to estimate the counting rates
since, while nuclear shadowing is theoretically important, it is expected 
to be less than factor of two for gluons in the $p_T$ range we discuss. 
%%%%
We calculate inclusive rates as a function of gluon $x$ and jet $p_T$ 
to compare with HERA.

The coherent diffractive processes we study are characterized by gaps in the 
directions of both nuclei and two jet production near the edge of the 
rapidity interval where hadrons are produced, as is the case for direct 
photon-nucleus/nucleon interactions. The diffractive rate is intimately 
related to the phenomenon of nuclear shadowing and the onset of the black disk 
limit where coherent diffraction would be 50\% of the total cross section. 
In our numerical studies, we used the only nuclear diffractive PDFs currently 
available \cite{Frankfurt:2003gx}, based on the leading twist  
description of hard diffraction in $ep$ scattering and using 
a quasi-eikonal approximation to model diffraction off more than 3 nucleons.  

In our calculations, we focus on small $x$ kinematics where a photon 
interacts predominantly with a gluon via photon-gluon fusion.  Since the
direct mechanism dominates in this kinematics, we ignore the partonic 
constituents of the photon.  We define $x_1$ as the momentum fraction 
carried by the photon
while  $x_2$ is the momentum fraction carried by the gluon from the nucleus 
(both per nucleon) or proton (see Fig.~\ref{1}).
The average jet rapidity is $y\sim -0.5 \ln(x_1/x_2)$. The photon $x_1$ is 
related to the jet $p_T$ by $x_1x_2\sim 4p_T^2/s_{NN}$.
 %--------------------------------------------
\begin{figure}[htb]
\includegraphics[width=8cm]{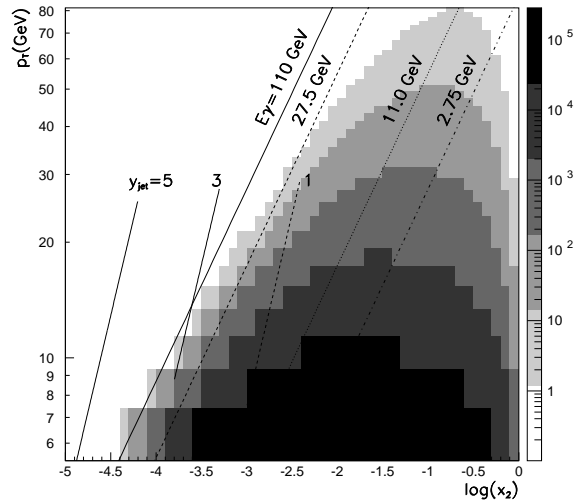}
\caption{The expected dijet photoproduction rate for a one month LHC Pb+Pb run
at $0.42 \times 10^{27}$cm$^{-2}$s$^{-1}$.
Rates are in counts per bin of $\pm 0.25  x_2$ and $\pm 1$ GeV in $p_T$.}
\label{2}
\end{figure}
%--------------------------------------------

Event rates were calculated for bins of $\pm 1$ GeV in $p_T$ and 
$\pm0.25  x_2$ in $x_2$ for a nominal one month run of $10^6$ seconds. 
The average luminosity is taken to be $0.42\times 10^{27}$cm$^{-2}$s$^{-1}$
for Pb+Pb collisions \cite{lum1} and $7.4\times10^{29}$cm$^{-2}$s$^{-1}$ for 
$p$Pb collisions \cite{lum2}. 
	
Ultraperipheral collision data can be recorded exploiting the full live time 
available as long as an acceptable trigger can be found.  The PHENIX
collaboration at RHIC found that a loose UPC trigger in Au+Au runs yielded a 
trigger rate of $<0.5\% \, \sigma_{\rm inel}$ ($10-20$ Hz at the LHC) 
\cite{PHENIX}. The PHENIX UPC $J/\psi$ trigger required a single high $p_T$ 
electron, a rapidity gap and a leading neutron tag. 
%--------------------------------------------
\begin{figure}[htb]
%\vspace{9pt}
\includegraphics[width=8cm]{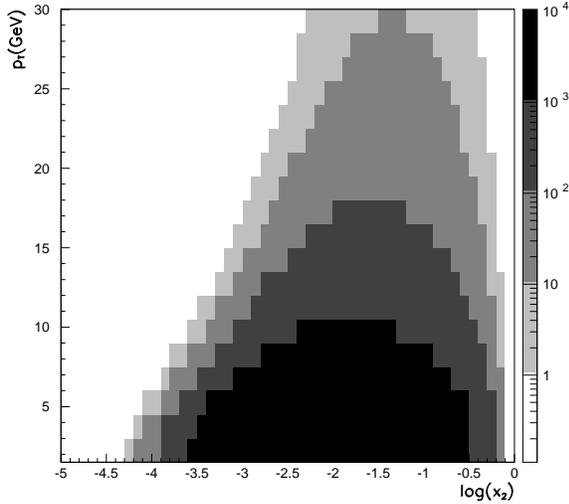}
\caption{The rate for $b$-jet photoproduction. Here the $p_T$
 bins are $\pm 0.75$ GeV.}
\label{3}
\end{figure}
%--------------------------------------------

In our case, there is always a high $p_T$ jet as well as the potential for
a heavy flavor tag using soft leptons or a secondary vertex, well within the 
calorimeter and tracking acceptance of ATLAS, for example. 
Because the ATLAS calorimeter extends $\pm 4.9$ units in $\eta$,
it should be possible to veto on activity with $E_T \geq 2$ GeV 
within $2.9<|\eta|<4.9$ along the direction of the ion emitting 
the exchanged photon. The target lead ion will always emit an evaporation 
neutron which can be tagged with the ATLAS Zero Degree Calorimeters except 
in the case of coherent diffractive photoproduction. Furthermore, the neutron 
multiplicity is correlated with
centrality in photon-nucleus collisions and hence can be used
to select central collisions where high gluon density effects are maximized.  

In Fig.~\ref{2} we present the counting rate per $x_2$ and 
$p_T$ bin accumulated in a one month LHC Pb+Pb run. 
In addition to the rates, 
we show contours of constant photon energy. 
Photon energies of a few GeV up
to $\sim 100$ GeV dominate the rates.  Under normal conditions, these photons
will collide with nuclei at energies of 2.75 TeV/nucleon. Contours of constant 
average jet rapidity are also shown in Fig.~\ref{2}. 
It is significant that jet production is predominantly in the region 
$\left|y_{\rm jet}\right|\le 3$ and therefore well matched 
to the calorimeter coverages of ATLAS and CMS at the LHC. The rapidity gap 
distinguishes these photoproduced jets from those produced in the 
more abundant nuclear collisions.
 %--------------------------------------------
\begin{figure}[htb]
%\vspace{9pt}
\includegraphics[width=8cm]{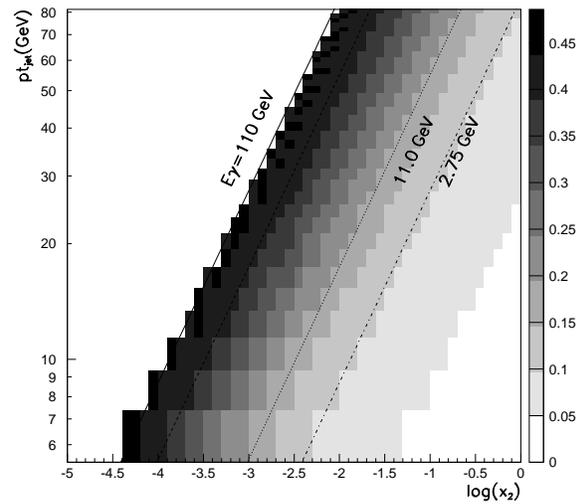}
\caption{The fraction of photoproduction events where 
additional photon exchanges result in the breakup of both beam nuclei.}
\label{4}
\end{figure}
%--------------------------------------------

The ATLAS experiment was designed to study jet production at $p_T> 10$ GeV. 
It is not yet clear what lower $p_T$ limit will apply to photoproduction 
where neither the ``underlying event"
nor multiple event pileup are present to degrade the measurement. 
The counting rate per bin, $100-1000$ for $p_T\ge 10-20$ GeV, should be
high enough for meaningful statistics.

Figure~\ref{3} shows the $b$-jet photoproduction counting rate.  These heavy
flavor jets are also photoproduced copiously at the LHC.  They can be 
detected via a detached vertex using the ATLAS pixel tracker or via a soft 
lepton tag.  We also calculated gamma-jet production but, since the rates 
are reduced by several orders of magnitude 
relative to the dijet rates, they are not shown.
	
We have also calculated diffractive photoproduction of dijets and $b$ jets 
employing the leading twist analysis of the diffractive PDFs in
Ref.~\cite{Frankfurt:2003gx}. 
This analysis suggests that even at $p_T \sim 5 -10$ GeV,  diffractive 
nuclear interactions at $x_2 \le 10^{-3}$ will be rather close to the black 
disk limit where more than 20\% of events are diffractive.  Accordingly, the 
diffractive rates are large over most of the $x_2$ range.

However, measurements of diffractive jet production present a triggering 
challenge for ATLAS since essentially the only activity in the event is a 
pair of soft jets at midrapidity. The target nucleus does not break up as 
in the non-diffractive case. Therefore, we have also calculated
photoproduction cross sections including the exchange of low energy photons.
These low energy photons are plentiful since the strength parameter, 
$Z^2 \alpha$, is of order unity in heavy ion collisions. 
The calculation proceeds as in Ref.~\cite{baur_et_al}. As seen in
Fig.~\ref{4}, the fraction of events with additional 
photon exchanges which break up both beam nuclei is strictly a 
function of primary photon energy and is $\sim 20-40\%$ over most 
of our kinematic range.
%--------------------------------------------
\begin{figure}[htb]
%\vspace{9pt}
\includegraphics[width=8cm]{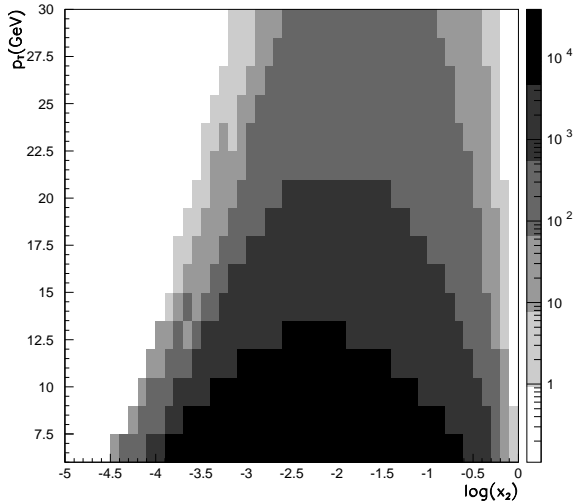}
\caption{The expected $b$-jet photoproduction rate in a one month LHC $p$Pb run
at  $7.4 \times 10^{29}$cm$^{-2}$s$^{-1}$.}
\label{5}
\end{figure}
%--------------------------------------------

The LHC can also run with asymmetric species so that $p$Pb collisions are 
likely to be a part of the heavy ion program. Under these conditions, the LHC 
will extend to higher energy, albeit with small $Q^2$, the 
proton structure measurements carried out at HERA. Figure~\ref{5} shows 
the $b$-jet photoproduction rates for collisions of photons emitted by the
lead nucleus with gluons from the proton beam.  The $b$-jet 
photoproduction rate is 
considerably higher than in Pb+Pb, primarily due to the increased energy of
the proton beam relative to the lead beam.
The charm jet rate is at least a factor of four larger.  
Heavy quark production is optimal in $pA$ since
the strong interaction contribution due to screened Pomeron exchange becomes 
important in dijet production \cite{Guzey:2005ys}.  The diffractive rates 
should also be a significant fraction of the total rate, $\ge 10\%$, 
allowing study of the diffractive proton PDFs at smaller $x$ than at HERA.
If the 420 m tagging proton stations \cite{loi} are implemented in the LHC, 
it will also be possible to considerably extend the HERA $t$-dependence 
measurements of the diffractive nucleon PDFs.

Our calculations have assumed that the linear DGLAP approximation is valid.
However, at the lowest values of $x_2$ and $p_T$ we study, DGLAP evolution
may break down.  The validity of the DGLAP approximation at a given $x$ and
$Q^2$ can be characterized by the ratio of the first nonlinear evolution term 
to the DGLAP linear term,
$C\alpha_s(Q^2)xg(x,Q^2)/Q^2R_T^2$, where $R_T$ is the radius of the target.  
The coefficient $C$ is a factor of 9/4 larger for processes dominated by  
direct coupling to gluons relative to quark couplings.  The lowest $x$ and 
$Q^2$ bin for $F_2(x,Q^2)$ measurements at HERA is at $x = 10^{-4}$ and $Q^2=
4$ GeV$^2$.  On the other hand, UPC measurements at the LHC can reach
$x \sim 5 \times 10^{-5}$ for our minimum $p_T$ of 5 GeV.  To quantify how
much further UPCs are into the nonlinear regime than are $ep$ collisions at
HERA, we form the double ratio \cite{Arneodo:1996qa},
\begin{equation}
 { (9/4) 0.7 \, A^{1/3} \alpha_s(p_{T}^2)xg(x\sim 5\times 10^{-5},p_T^2)/p_T^2 
\over  \alpha_s(Q^2)xg( x\sim 10^{-4},Q^2)/Q^2}\sim  3 \,\, . \label{ratrat}
  \end{equation}
The factor of $0.7$ is the ratio $(r_N/R_A)^2$ multiplied by a factor of
1.5 for photons going through the center of the nucleus.
To calculate Eq.~(\ref{ratrat}), we used recent leading-order parton density 
fits and neglected leading-twist nuclear shadowing, a small correction for 
$p_T \geq 5$ GeV.  
Note that by comparing the ratios at the same values of $x$ and $Q^2$,
Eq.~(\ref{ratrat}) becomes $(9/4) 0.7 \, A^{1/3} \sim 9$ for lead beams. 
We can form a similar ratio of UPCs at the LHC to $eA$ collisions 
at the proposed eRHIC
collider where the lowest $x$ is $\sim 10^{-3}$ for $Q^2\sim 4$ GeV$^2$, 
finding a relative ratio of 
$\sim 1.5$.  The increase in the importance of the nonlinear regime for UPCs
at the LHC relative to $ep$ and $eA$ collisions are due to the direct gluon 
coupling and a much larger $x$ range as well as the use of nuclear beams 
(relative to $ep$ collisions).

We thus demonstrate that UPCs probe the nuclear PDFs at $p_T\ge 5$ GeV 
over an $x$ range a factor of ten greater than at HERA as well as
determine the nuclear diffractive PDFs in much of the same $x$ range. 
An LHC $pA$ run will also extend the HERA $x$ range of the inclusive
and diffractive 
gluon PDFs in the proton a factor of ten.
 
{\it Acknowledgments}: We would like to thank V.~Guzey for helpful 
discussions. This work was supported in part by the US Department of
Energy, Contract Numbers DE-AC02-05CH11231 (R. Vogt); 
DE-FG02-93ER40771 (M. Strikman); and DE-AC02-98CH10886 (S.N. White).

\end{document}